\documentclass[journal,twoside,web]{IEEEtran}
\usepackage{generic}
\usepackage{cite}
\usepackage{amsmath,amssymb,amsfonts}
\usepackage{algorithmic}
\usepackage{graphicx}
\usepackage{algorithm,algorithmic}
\usepackage{hyperref}
\hypersetup{hidelinks=true}
\usepackage{textcomp}
\usepackage{booktabs} 
\usepackage{pifont}
\usepackage{xcolor}
\usepackage{graphicx}
\usepackage{balance}
\usepackage[linesnumbered,ruled,vlined]{}

\def\BibTeX{{\rm B\kern-.05em{\sc i\kern-.025em b}\kern-.08em
    T\kern-.1667em\lower.7ex\hbox{E}\kern-.125emX}}
\begin{document}
\title{T-CACE: A Time-Conditioned Autoregressive Contrast Enhancement Multi-Task Framework for Contrast-Free Liver MRI Synthesis, Segmentation, and Diagnosis}
\author{Xiaojiao Xiao, Jianfeng Zhao, Qinmin Vivian Hu, and Guanghui Wang, \IEEEmembership{Senior Member, IEEE}
\thanks{This work is partly supported by the Natural Sciences and Engineering Research Council of Canada (NSERC) and TMU FOS Postdoctoral Fellowship.}
\thanks{X. Xiao, Q. Hu, and G. Wang are with the Department of Computer Science, Toronto Metropolitan University, Toronto, Canada (e-mail:\{xiaojiao, vivian, wangcs\}@torontomu.ca)}
\thanks{J. Zhao is with the School of Biomedical Engineering, Western University, London, Canada (e-mail:jzhao525@uwo.ca)}
}

\maketitle

\begin{abstract}
Magnetic resonance imaging (MRI) is a leading modality for the diagnosis of liver cancer, significantly improving the classification of the lesion and patient outcomes. However, traditional MRI faces challenges including risks from contrast agent (CA) administration, time-consuming manual assessment, and limited annotated datasets. To address these limitations, we propose a Time-Conditioned Autoregressive Contrast Enhancement (T-CACE) framework for synthesizing multi-phase contrast-enhanced MRI (CEMRI) directly from non-contrast MRI (NCMRI). T-CACE introduces three core innovations: a conditional token encoding (CTE) mechanism that unifies anatomical priors and temporal phase information into latent representations; and a dynamic time-aware attention mask (DTAM) that adaptively modulates inter-phase information flow using a Gaussian-decayed attention mechanism, ensuring smooth and physiologically plausible transitions across phases. Furthermore, a constraint for temporal classification consistency (TCC) aligns the lesion classification output with the evolution of the physiological signal, further enhancing diagnostic reliability. Extensive experiments on two independent liver MRI datasets demonstrate that T-CACE outperforms state-of-the-art methods in image synthesis, segmentation, and lesion classification. This framework offers a clinically relevant and efficient alternative to traditional contrast-enhanced imaging, improving safety, diagnostic efficiency, and reliability for the assessment of liver lesion. The implementation of T-CACE is publicly available at: https://github.com/xiaojiao929/T-CACE. 

\begin{IEEEkeywords}
Autoregressive model; MRI synthesis; Liver tumor classification; Non-contrast MRI; Segmentation
\end{IEEEkeywords}


\end{abstract}

\section{Introduction}
\label{sec:introduction}

\begin{figure}[t]
	\centering
	\includegraphics[width=0.49\textwidth]{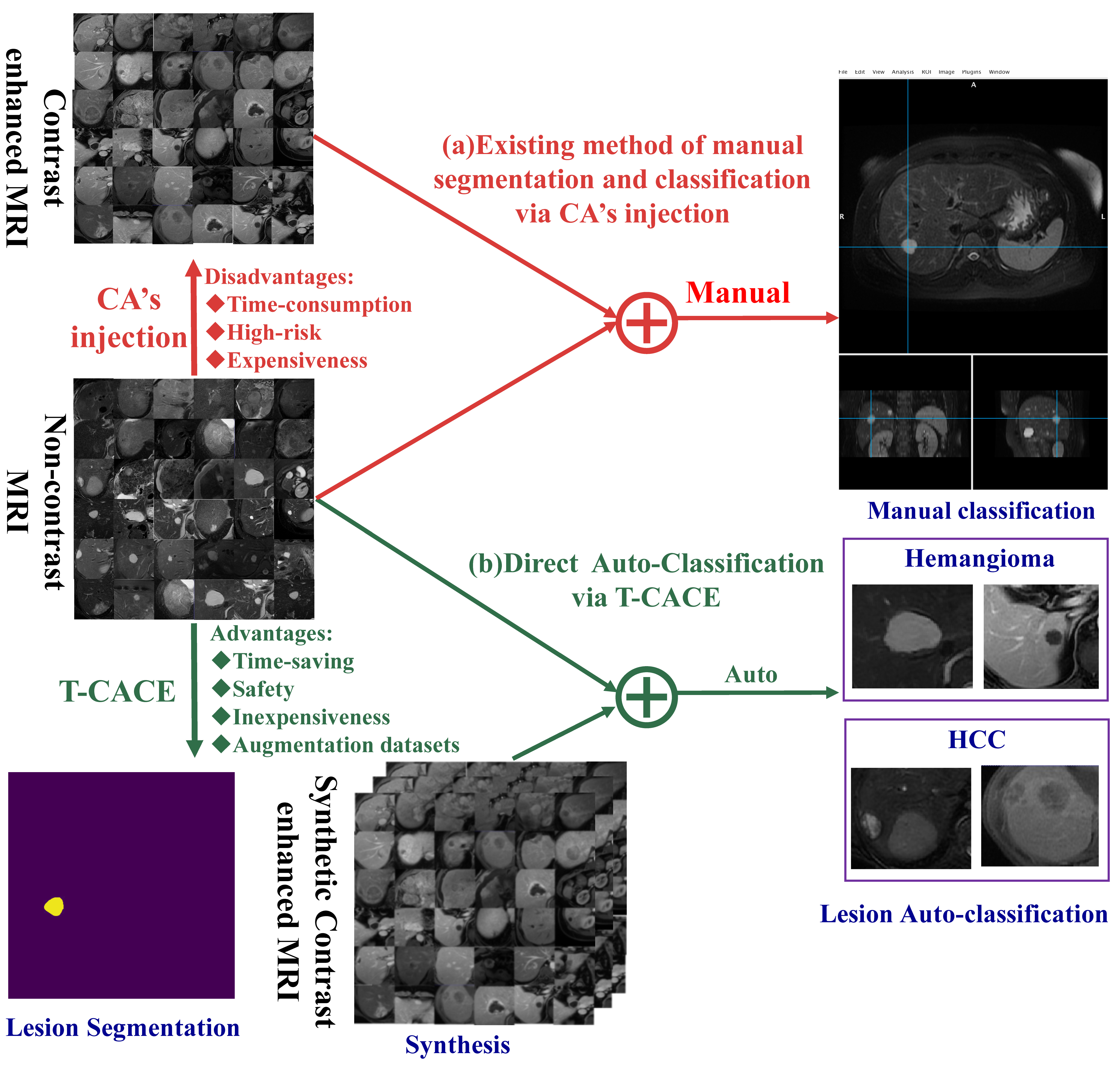}
	\caption{Motivation and overview of the proposed framework. (a) Conventional liver lesion diagnosis relies on manual segmentation and classification using contrast-enhanced MRI, which requires contrast agent administration and is subject to inter-observer variability. (b) In contrast, the proposed T-CACE framework enables fully automated, contrast-free synthesis, segmentation, and classification directly from non-contrast MRI. By unifying these tasks within a time-conditioned autoregressive architecture, T-CACE improves diagnostic efficiency, consistency, and safety.} \label{figure1}
\end{figure}

\IEEEPARstart{L}{iver} cancer remains one of the leading causes of cancer-related mortality worldwide, posing a substantial public health burden \cite{Ferlay2010MICC}. Contrast-enhanced magnetic resonance imaging (CEMRI) plays a pivotal role in the diagnosis of liver disease by providing high-resolution soft tissue contrast and enabling accurate differentiation between benign and malignant lesions \cite{Semelka2001}. However, reliance on CEMRI introduces notable challenges, including labor-intensive manual interpretation, variability between radiologists, and the need for contrast agent injection (CA), increasing examination cost, duration, and risk, particularly for patients with renal impairment \cite{Marckmann2006}. Consequently, there is a pressing clinical need for alternative approaches that enable accurate segmentation and classification of the lesion without CA administration, as illustrated in Fig.~\ref{figure1}.

Recently, researchers have explored non-contrast MRI (NCMRI) as a promising diagnostic tool, leveraging deep learning to bridge the contrast gap between NCMRI and CEMRI. Wu et al. \cite{wu2019radiomics} proposed an automatic radiomics-based classifier to distinguish hepatocellular carcinoma (HCC) from hemangiomas using NCMRI. Xu et al. \cite{xu2021synthesis} proposed a pixel-level graph reinforcement learning framework to synthesize contrast-enhanced MRI from non-contrast inputs. However, their method processes each phase independently and lacks explicit temporal modeling across arterial, portal venous, and delayed phases. This may compromise diagnostic accuracy, especially for tasks relying on dynamic signal patterns such as lesion classification. Zhao et al. \cite{zhao2020tripartite} introduced Tripartite-GAN to generate multi-phase CEMRI for tumor detection. While effective in image synthesis, their framework handles synthesis and detection separately, without joint optimization. This disconnect can lead to task inconsistency, where synthesized features may not align well with classification needs, thereby limiting diagnostic performance.

 Multi-phase CEMRI is especially valuable in liver imaging, as it captures dynamic lesion enhancement across arterial (Art), portal venous (PV), and delayed (Delay) phases. Each phase provides distinct diagnostic cues—hypervascular tumor detection in ART, improved lesion-parenchyma contrast in PV, and contrast washout in Delay, which is a key indicator of malignancy \cite{kim2019hepatobiliary}. These enhancement characteristics are essential for clinical decision-making, making multi-phase synthesis more informative than single-phase generation. However, existing methods often overlook temporal consistency in multi-phase enhancement and struggle to propagate phase-specific features, resulting in unrealistic contrast transitions and inconsistent lesion enhancement.

 Synthesizing multi-phase CEMRI from NCMRI presents three major challenges. First, there is inconsistency across multi-task predictions: the optimization objectives of synthesis, segmentation, and classification tasks differ, leading to mismatches in the prediction spaces. For example, the synthesized CEMRI may show contrast distribution shifts in specific regions, hindering accurate lesion segmentation, while uncertainty in lesion boundaries can further affect classification. Such task-level inconsistency results in unstable loss optimization and reduced diagnostic reliability. Second, cross-domain translation is inherently challenging: the mapping from NCMRI to CEMRI is highly nonlinear and physiologically dependent. Direct pixel-wise transformations or fixed regression models struggle to approximate this high-dimensional relationship, causing the synthesized images to deviate from real CEMRI distributions, which reduces clinical interpretability. Third, most existing methods inadequately model dynamic enhancement patterns: CEMRI comprises multiple phases with signal intensity evolving according to characteristic dynamic enhancement processes, which are crucial for reliable lesion interpretation and malignancy identification \cite{liu2013quantitatively}. Without temporal consistency modeling, the synthesized contrast trajectories may deviate from true lesion dynamics, undermining both lesion classification reliability and diagnostic stability.

To address these issues, we propose a novel Time-Conditioned Autoregressive Contrast Enhancement (T-CACE) framework for joint synthesis, segmentation, and classification of multi-phase liver MRI. Our goal is to enable accurate and consistent diagnostic modeling directly from non-contrast MRI (NCMRI), thereby offering a contrast-free alternative to conventional contrast-enhanced imaging. Specifically, to ensure consistency across tasks, we adopt a unified autoregressive strategy, where synthesized Art, PV, and Delay phase images are generated progressively, preserving the structural alignment between phases. To address domain gaps in NCMRI-to-CEMRI translation, we introduce Conditional Token Encoding (CTE) that integrates anatomical priors and temporal phase information into latent representations. A Temporal Consistency Constraint (TCC) is introduced to align the classification output with the physiological signal evolution across phases. Additionally, we design a Dynamic Time-aware Attention Mask (DTAM) that adaptively modulates inter-phase information flow using a Gaussian-decayed attention mechanism. This enables the model to emphasize temporally relevant features and achieve coherent multi-phase synthesis and segmentation.

The paper makes the following key contributions:

\begin{enumerate}
\item {\bf Time-Conditioned Autoregressive Contrast Enhancement (T-CACE)}: We propose a novel unified framework for the joint synthesis, segmentation, and classification of NCMRI. By incorporating time as a conditioning variable, T-CACE sequentially generates contrast-enhanced phases and aligns predictions from multiple tasks within a shared latent space, thereby mitigating multi-task inconsistency.

\item {\bf Conditional Token Encoding (CTE):}
To bridge the domain gap in NCMRI-to-CEMRI translation, we introduce a Conditional Token Encoding mechanism that integrates anatomical priors, temporal phase indicators, and continuous time encodings into unified latent tokens, capturing both spatial and temporal context.

\item {\bf Dynamic Time-aware Attention Mask (DTAM):}  
We further design a Dynamic Time-aware Attention Mask in the autoregressive module, employing a Gaussian-decayed attention mechanism to adaptively emphasize temporally relevant features and ensure smooth, physiologically coherent transitions across synthesized phases.

\item {\bf Temporal Classification Consistency (TCC):}  
To further enhance diagnostic reliability, we propose a Temporal Classification Consistency constraint that aligns lesion classification predictions with the physiological signal evolution derived from the synthesized phases. 
\end{enumerate}

\section{Related Work}

\subsection{Contrast-Enhanced Liver MRI Synthesis from Non-Contrast Data}

Synthesizing contrast-enhanced liver images from non-contrast MRI (NCMRI) has attracted increasing attention as a safer and more accessible alternative to contrast-based imaging \cite{xiao2024fgc2f}. Numerous studies have leveraged generative models, particularly GAN-based frameworks, to synthesize multi-phase contrast-enhanced MRI (CEMRI) with enhanced realism and clinical utility. Jiao et al. \cite{jiao2023contrast} introduced a sparse attention fusion GAN with gradient regularization for CEMRI synthesis, significantly improving perceptual fidelity and lesion visibility. Wang et al. \cite{wang2021contrast} proposed an attention-guided CycleGAN to translate NCMRI into contrast-enhanced domains, capturing key enhancement patterns relevant to diagnosis. In addition, Xu et al. \cite{xu2021synthesis} designed a pixel-level graph reinforcement learning network to simulate gadolinium-based enhancement in liver MRIs, thereby facilitating contrast-free tumor analysis.

Beyond MRI, similar approaches have been adopted in contrast-enhanced CT (CECT) synthesis. Song et al. \cite{song2020non} employed CycleGAN-based augmentation to generate synthetic CECT from non-contrast CT, improving segmentation accuracy on liver structures. Zhong et al. \cite{zhong2023united} proposed a unified multi-task learning framework that jointly performs deformable registration and CECT synthesis, enhancing anatomical consistency and cross-domain fidelity. These synthetic frameworks not only reduce dependency on contrast agents but also benefit downstream learning tasks. For example, Xu et al. \cite{xu2024accurate} demonstrated that synthetic multi-modal data significantly boosts liver tumor segmentation performance, while Frid-Adar et al. \cite{frid2018gan} showed that GAN-generated images can improve CNN-based liver lesion classification. Although these methods have advanced contrast-free synthesis, challenges persist in accurately capturing physiological enhancement dynamics and ensuring consistent signal evolution across multiple contrast phases.

\subsection{Autoregressive Modeling for Medical Image Synthesis and Analysis}

Autoregressive models have garnered increasing attention in medical imaging for their ability to model sequential dependencies and complex distributions, building upon early successes in natural language processing \cite{chen2023evoprompting}. Motivated by these advances, researchers have adapted autoregressive frameworks to a wide range of imaging tasks and anatomical sites. For instance, Wang et al. \cite{wang2024autoregressive} developed an autoregressive pre-training framework for 3D medical image representation, significantly enhancing downstream performance in tasks including lesion detection and segmentation. Similarly, Tudosiu et al. \cite{tudosiu2022morphology} introduced a morphology-preserving autoregressive generative model, enabling the synthesis of anatomically accurate, high-resolution brain MR images, thus facilitating reliable clinical decision-making and surgical planning. Furthermore, Gui et al. \cite{gui2024cavm} proposed the Conditional Autoregressive Vision Model (CAVM), progressively generating contrast-enhanced MR images from non-contrast scans, which substantially improved image realism and clinical interpretability by preserving the temporal consistency inherent in multi-phase imaging.

Autoregressive modeling has also been successfully applied to medical time-series data. Li et al. \cite{li2023causal} developed a causal recurrent variational autoencoder (CR-VAE) for generating temporally dependent longitudinal clinical data. In image reconstruction, Kabas et al. \cite{kabas2024physics} integrated imaging physics into an autoregressive state-space model, markedly improving reconstruction from undersampled data. For segmentation, Zhang et al. \cite{chen2025autoregressive} presented a next-scale mask prediction framework based on autoregressive principles, sequentially refining segmentation accuracy across multiple scales. Collectively, these works demonstrate the versatility and impact of autoregressive approaches in advancing medical image synthesis, reconstruction, and segmentation.

\subsection{Deep Learning Methods for Liver Lesion Classification}
Accurate classification of liver lesions is critical for early diagnosis and treatment planning. Recent studies have explored deep learning and hybrid strategies to distinguish lesion types using both contrast-enhanced and non-contrast imaging. Yasaka et al. \cite{yasaka2018deep} proposed a CNN-based framework achieving 84\% accuracy in differentiating liver lesion types from CEMRI. Trivizakis et al. \cite{trivizakis2018extending} designed a 3D convolutional model that utilizes inter-slice context to distinguish between primary and metastatic liver tumors. To enhance robustness, several works have explored data augmentation and synthesis for classification. Frid-Adar et al. \cite{frid2018} demonstrated that GAN-based image generation can significantly improve CNN performance in liver lesion classification. Wu et al. \cite{wu2019radiomics} used radiomics features derived from NCMRI to classify hepatocellular carcinoma and hemangiomas, demonstrating the feasibility of contrast-free classification. Wang et al. \cite{wang2024mix} proposed a Mix-Domain Contrastive Learning (MDCL) for unpaired H\&E-to-IHC stain translation. Zhao et al. \cite{zhao2020tripartite} developed a Tripartite-GAN to synthesize contrast-enhanced MR phases, thereby improving tumor detection by capturing morphological enhancement characteristics. These approaches support the growing trend of integrating image synthesis with lesion classification, particularly in contrast-free settings, where preserving enhancement dynamics is essential for stable and reliable diagnostic outcomes.

\section{Methodology}
\subsection{Problem Formulation and Method Overview}
Given a non-contrast liver MRI scan \( x_{NC} \) and its corresponding tumor mask \( x_{TM} \), our model simultaneously synthesizes three clinically relevant phases of contrast-enhanced MRI (CEMRI): arterial phase \( y_{Art} \), portal venous phase \( y_{PV} \), and delayed phase \( y_{Delay} \), without the administration of contrast agents (CA). In addition, the model performs lesion segmentation \( \hat{y}^{seg} \) and lesion classification \( y^{\text{cls}} \), thus providing comprehensive diagnostic support. 

\begin{figure}[t]
	\centering
	\includegraphics[width=0.49\textwidth]{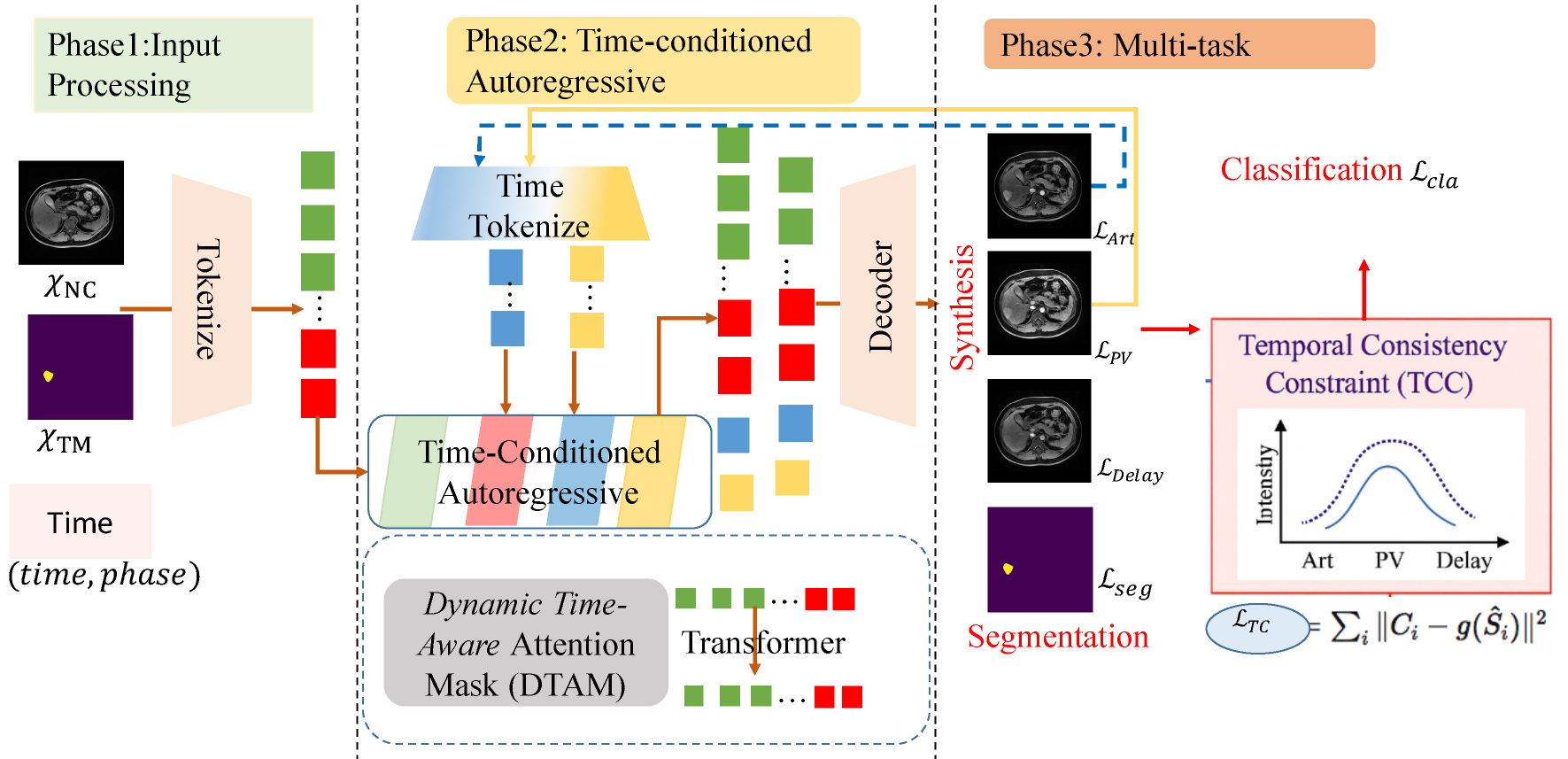}
	\caption{Overall architecture of the proposed T-CACE framework for multi-task liver MRI synthesis, segmentation, and classification.
The framework comprises three stages: (1) Input Processing, where the non-contrast MRI, tumor mask, and continuous time annotations are tokenized into conditional tokens and time-phase tokens; (2) Time-Conditioned Token Encoding, where a transformer equipped with Conditional Token Encoding (CTE) and Dynamic Time-Aware Attention Mask (DTAM) models temporal dependencies across phases in an autoregressive manner; and (3) Multi-task Decoding, where dedicated decoders simultaneously perform contrast-enhanced MRI synthesis, lesion segmentation, and malignancy classification. A Temporal Consistency Constraint (TCC) aligns image-derived and signal-derived classification outputs, improving the clinical plausibility and diagnostic consistency of the generated results. } \label{method1}
\end{figure}

We address this multi-task challenge as a sequential prediction problem, progressively modeling contrast enhancement using a time-conditioned autoregressive approach explicitly constrained by temporal dependencies. Our proposed framework, detailed in Fig.~\ref{method1}, involves several steps: first, the inputs \( (x_{NC}, x_{TM}) \) undergo conditional token encoding (CTE), incorporating anatomical priors and temporal information. To explicitly encode temporal phase information (Art, PV, Delay), we introduce both discrete phase tokens \( t_{\mathrm{phase}} \), created via a learnable embedding from the discrete phase labels, and continuous time encoding tokens \( t_{\mathrm{time}} \), constructed as sinusoidal functions of normalized phase acquisition times. Subsequently, the combined tokens, including phase-specific tokens, are processed through our novel Dynamic Time-aware Attention Mask (DTAM) within the autoregressive module, ensuring temporal coherence and precise modeling of contrast dynamics. Finally, a hierarchical transformer-based decoder reconstructs the synthesized multi-phase MRI outputs, concurrently predicting the lesion segmentation masks and lesion classifications, maintaining anatomical accuracy, physiological plausibility, and diagnostic relevance across the synthesized phases.

\subsection{Model Architecture}

\subsubsection{Conditional Token and Phase-aware Embedding}
In multi-phase contrast-enhanced MRI (CEMRI) synthesis, ensuring structural consistency across different phases while preserving temporal dynamics is crucial. Directly processing raw input images (i.e., \( x_{NC}, x_{TM} \)) independently at each phase may introduce inconsistencies in lesion structure and intensity distribution. To mitigate this, we introduce a Conditional Token Encoding (CTE) mechanism to effectively capture both spatial structure and temporal priors.

The original token \( t_{OT} \) is computed from the inputs \( x_{NC} \) and \( x_{TM} \) using a hierarchical feature encoder, $t_{OT} = W_{\text{proj}} f_{\text{enc}}(x_{NC}, x_{TM})$. \( f_{\text{enc}}(\cdot) \) is a deep feature extractor. Specifically, we adopt Swin UNETR \cite{cao2022swin}, a hierarchical vision transformer designed for medical imaging tasks. It extracts a compact latent representation by encoding spatial and anatomical details while maintaining global dependencies. The final layer of Swin UNETR is used to obtain the latent tokens \( t_{OT} \).

We introduce an additional discrete temporal phase token \( t_{\text{phase}}^{i} \), which explicitly encodes the specific contrast phase being synthesized (Art, PV, and Delay). A learnable embedding generates this token: $t_{\text{phase}}^{i} = \mathrm{Embedding}(\text{phase}_{i}), \quad \text{phase}_{i} \in \{\text{Art},\, \text{PV},\, \text{Delay}\}$. Furthermore, to accurately represent the temporal aspect associated with each synthesized contrast phase, we include a continuous time encoding token $ t_{\text{time}}^{i} = [\sin(\omega t_{i}),\, \cos(\omega t_{i})]$, 
where \( t \) represents the elapsed time since the hypothetical contrast agent administration, and \( \omega \) is a frequency modulation parameter ensuring smooth and continuous temporal transitions.

The final conditional token for phase $i$ is structured as $t_{\text{CT}^{i}} = [t_{OT}, t_{phase}^{i}, t_{\text{time}}^{i}]$. This integrated representation effectively conditions the autoregressive synthesis process on anatomical consistency, explicit contrast-phase identification, and continuous temporal evolution, ensuring coherent and physiologically plausible synthesis results.

\subsubsection{Time-Conditioned Autoregressive Module with DTAM}
To guarantee temporal coherence and smooth physiological transitions between synthesized phases, we employ a time-conditioned autoregressive framework augmented with a Dynamic Time-aware Attention Mask (DTAM). 

\begin{figure}[t]
	\centering
	\includegraphics[width=0.49\textwidth]{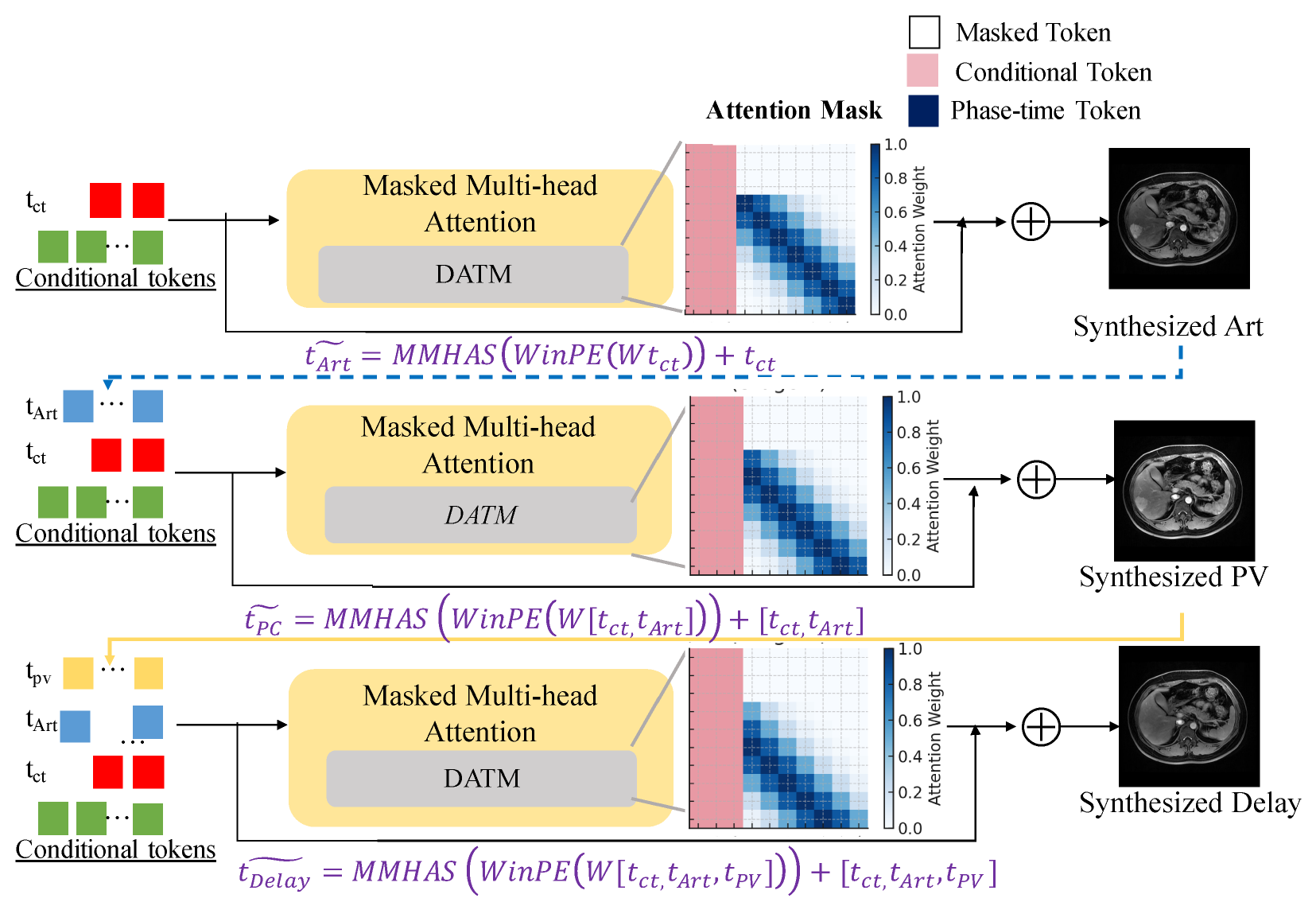}
	\caption{Autoregressive multi-phase CEMRI synthesis pipeline based on Dynamic Time-Aware Attention Masking (DTAM). This figure illustrates the DTAM-guided autoregressive synthesis process for three CEMRI phases (Arterial, Portal Venous, and Delayed). At each stage, the left pink-shaded region denotes fixed conditional tokens (e.g., non-contrast images and tumor masks), which remain constant throughout the process. The right blue-shaded region contains phase-specific latent tokens, progressively updated autoregressively. The attention maps visualize dynamic weighting: attention concentrates along the temporal diagonal while adaptively attending to relevant prior tokens. This mechanism enables phase-aware feature fusion and ensures temporally coherent synthesis across the multi-phase CEMRI pipeline. } \label{method2}
\end{figure}

\textbf{Dynamic Time-aware Attention Mask (DTAM):} At the initial step, the network receives the conditional token $t_{\text{CT}}$, which encodes the anatomical features and temporal priors specific to the arterial phase. The first-phase image $y_{\text{Art}}$ is synthesized solely based on this conditional token. For subsequent phases, such as PV and Delay, the synthesis becomes autoregressive: the model not only receives the corresponding conditional token $t_{\text{CT}}^{i}$ (for $i = \text{Art}, \text{PV}, \text{Delay}$), but also integrates information from the tokens and features derived from all previously generated phase images (e.g., the token embedding of $t_{\text{Art}}$ for PV, and both $t_{\text{Art}}$ and $t_{\text{PV}}$ for Delay), as shown in Fig.\ref{method2}.

Formally, for the $i$-th phase, the input to the dynamic attention mechanism consists of the current conditional token $t_{\text{CT}}^{i}$ as well as the set of previously generated image tokens $\{t_{1}, t_{2}, ..., t_{i-1}\}$, each embedded into the same latent space. The attention output $z_i$ is then computed as a weighted sum:
\begin{equation}
z_i = \sum_{k=1}^{i} a_{ik} v_k,
\end{equation}
where $v_k$ are value embeddings of either the initial conditional token or the previously synthesized phase tokens. The dynamic attention weight $a_{ik}$ is defined as:
\begin{equation}
a_{ik} = 
G(i, k) \cdot \frac{\exp(Q_i K_k^\top)}{\sum_{j=1}^{i} G(i, j) \exp(Q_i K_j^\top)},
\end{equation}
where the Gaussian decay function $G(i, k)$ ensures temporally adjacent phases have greater influence:
\begin{equation}
G(i, k) = \exp \left(-\frac{(t_i - t_k)^2}{2\sigma^2}\right).
\end{equation}

Here, $Q_i$ and $K_k$ are the query and key projections of the $i$-th and $k$-th tokens (including both conditional tokens and previous image embeddings), $t_i$ and $t_k$ are their corresponding time encodings, and $\sigma$ controls the temporal decay rate. The $\sigma$ is empirically set to 0.7, which was selected based on a sensitivity analysis for achieving optimal cross-phase interaction while avoiding over-smoothing.

This dynamic masking approach ensures that, at each phase, the model adaptively attends to both anatomical priors and the most temporally relevant contextual features from already synthesized phases. The autoregressive design, reinforced by DTAM, allows the network to synthesize each phase in strict sequence, leveraging local detail and global enhancement trends for smooth, realistic CEMRI progression.

\begin{algorithm}[t]
\caption{Time-Conditioned Autoregressive Multi-task Framework}
\begin{algorithmic}[1]
\STATE \textbf{Input:} Non-contrast MRI $x_{NC}$, tumor mask $x_{TM}$
\STATE \textbf{Output:} Synthesized phases $\{ \hat{y}_{Art}, \hat{y}_{PV}, \hat{y}_{Delay} \}$, aggregated segmentation mask $\hat{y}^{seg}$, classification output $\hat{y}^{cls}$

\vspace{1mm}
\STATE \textbf{1. Conditional Token Encoding and Embedding:}
\STATE $t_{OT} \leftarrow W_{\text{proj}} f_{\text{enc}}(x_{NC}, x_{TM})$ \hfill \textit{// Organ-level token}
\STATE $t_{\text{time}} \leftarrow [\sin(\omega t), \cos(\omega t)]$ \hfill \textit{// Continuous time embedding}
\STATE $t_{\text{phase}} \leftarrow \text{Phase Index Embedding}$ 
\STATE $t_{CT} \leftarrow [t_{OT}, t_{\text{phase}}, t_{\text{time}}]$ \hfill \textit{// Conditional tokens (shared across phases)}

\vspace{1mm}
\STATE \textbf{2. Autoregressive Synthesis with DTAM:}
\FOR{$i = 1$ to $3$ (Art, PV, Delay)}
    \IF{$i == 1$}
        \STATE $t_{\text{input}} \leftarrow t_{CT}$
    \ELSE
        \STATE $t_{\text{input}} \leftarrow [t_{CT}, \hat{t}_{1:i-1}]$ \hfill \textit{// Append prior phase tokens}
    \ENDIF
    \STATE $z_i \leftarrow \text{MMHSA}(\text{WinPE}(t_{\text{input}})) + t_{\text{input}}$
    \STATE $[\hat{y}_i, s_i, f_i] \leftarrow f_D(z_i)$ \hfill \textit{// Synthesis, segmentation, features}
    \STATE Store $s_i$ and $f_i$ for later aggregation
\ENDFOR
\STATE $\hat{y}^{seg} \leftarrow \mathcal{A}(s_{Art}, s_{PV}, s_{Delay})$

\vspace{1mm}
\STATE \textbf{3. Classification via Feature Fusion:}
\STATE $\mathbf{F}_{\text{joint}} \leftarrow \text{Concat}(f_{\text{feat}}(x_{NC}), f_{Art}, f_{PV}, f_{Delay})$
\STATE $\hat{y}^{cls} \leftarrow f_{\text{cls}}(\mathbf{F}_{\text{joint}})$

\vspace{1mm}
\STATE \textbf{4. Temporal Consistency Constraint (TCC):}
\STATE $\mathcal{L}_{TCC} = 0$
\FOR{each phase $i$}
    \STATE $\hat{S}_i \leftarrow f_{\text{signal}}(x_{NC}, t_i)$
    \STATE $label^{signal}_i \leftarrow g(\hat{S}_i)$
    \STATE $\mathcal{L}_{TCC} \leftarrow \mathcal{L}_{TCC} + \| \hat{y}^{cls}_i - label^{signal}_i \|^2$
\ENDFOR

\vspace{1mm}
\STATE \textbf{5. Multi-task Loss Optimization:}
\STATE $\mathcal{L}_{syn} = \sum_{i=1}^{3} \| \hat{y}_i - y^{GT}_i \|_1$
\STATE $\mathcal{L}_{seg} = \lambda_{\text{dice}} \mathcal{L}_{Dice}(\hat{y}^{seg}, y^{seg}) + \lambda_{\text{ce}} \mathcal{L}_{CE}(\hat{y}^{seg}, y^{seg})$
\STATE $\mathcal{L}_{cls} = \mathcal{L}_{CE}(\hat{y}^{cls}, y^{cls}_{GT})$
\STATE $\mathcal{L}_{total} = \mathcal{L}_{syn} + \mathcal{L}_{seg} + \mathcal{L}_{cls} + \mathcal{L}_{TCC}$
\STATE Update model parameters to minimize $\mathcal{L}_{total}$
\end{algorithmic}
\end{algorithm}

\subsubsection{Decoder for Multi-phase Synthesis and Segmentation}

We construct the autoregressive synthesis module based on a time-conditioned Transformer architecture, where the Dynamic Time-aware Attention Mask (DTAM) is integrated into each block to aggregate temporal context from previous phases. For each target phase, the model assembles the conditional token $t_{\text{CT}}^{i}$ together with embeddings of all previously generated images as input to the masked multi-head self-attention (MMHSA), which uses the window-based position encoding from Swin Transformer to maintain spatial and temporal awareness.

Mathematically, for $i \in \{\text{Art}, \text{PV}, \text{Delay}\}$, the autoregressive step is:
\begin{align}
    \tilde{t}_i = \text{MMHSA}\left(\text{WinPE}\left(W \cdot [t_{\text{CT}}^{i},\, \{\tilde{t}_k\}_{k=1}^{i-1}]\right)\right) + [t_{\text{CT}}^{i},\, \{\tilde{t}_k\}_{k=1}^{i-1}]
\end{align}
where $\text{WinPE}(\cdot)$ denotes the Swin window-based position encoding and $W$ is a linear projection.

The updated token $\tilde{t}_i$ is then decoded to generate the synthesized image and segmentation mask for phase $i$:
\begin{equation}
    [y_i,\, s_i] = f_D(\tilde{t}_i)
\end{equation}

To obtain the final lesion segmentation mask $\hat{y}^{seg}$, we aggregate the phase-wise predicted masks $\{s_{\text{Art}}, s_{\text{PV}}, s_{\text{Delay}}\}$ using majority voting. Specifically, for each pixel, the final segmentation label is determined by the most frequently predicted class among all three phases. This strategy effectively leverages the complementary information from different enhancement phases, improves segmentation robustness, and mitigates errors from any single phase prediction.

\subsubsection{Image-based and TCC-constrained Classification}

To classify lesions, we jointly utilize features from both the synthesized multi-phase images and the original non-contrast MRI (NCMRI). Specifically, the fused feature representation is defined as, $\mathbf{F}_{\mathrm{joint}} = \mathrm{Concat}(f_{\mathrm{feat}}(x_{NC}), f_{\mathrm{feat}}([\hat{y}_{Art}, \hat{y}_{PV}, \hat{y}_{Delay}])),$ where $f_{\mathrm{feat}}(\cdot)$ denotes the feature extraction module. The lesion classification prediction is then obtained by inputting the fused features into a classification head $\hat{y}^{\mathrm{cls}} = f_{\mathrm{cls}}(\mathbf{F}_{\mathrm{joint}}).$
    
To further ensure that the model’s lesion classification predictions are physiologically consistent with the temporal dynamics of contrast enhancement, we introduce a Temporal Consistency Constraint (TCC). Specifically, we define $\hat{S}_i$ as the predicted lesion enhancement intensity at temporal phase $i$. The prediction $\hat{S}_i$ is obtained by passing the non-contrast MRI $x_{NC}$ and a continuous, phase-specific time embedding $t_i \in \mathbb{R}^{2}$ into a dedicated neural network $f_{\theta}$, $ \hat{S}_i = f_{\theta}(x_{NC}, t_i).$

We clarify that the neural network $f_{\theta}(\cdot)$ is implemented as a lightweight 3-layer fully connected network. It takes as input a concatenation of the flattened latent feature vector extracted from the non-contrast MRI $x_{NC}$ (dimension $C=256$) and the continuous time embedding $t_i$, resulting in an input vector of shape $\mathbb{R}^{C+2}$. The network outputs a scalar $\hat{S}_i \in \mathbb{R}$ representing the predicted signal intensity at each temporal phase.

Next, the mapping function $g(\cdot)$ explicitly defines a threshold-based labeling rule to convert the predicted enhancement intensity $\hat{S}_i$ into a binary, physiologically meaningful diagnostic label:
\begin{equation}
    \mathrm{label}^{\mathrm{signal}}_i = g(\hat{S}_i) = \mathbb{I}(\hat{S}_i > \tau),
\end{equation}
where $\tau$ is a clinically relevant threshold indicative of lesion malignancy, and the indicator function $\mathbb{I}(\cdot)$ outputs binary labels corresponding to the "washout" enhancement pattern. In practice, we empirically set the threshold $\tau = 0.5$, which corresponds to the mid-range normalized signal intensity used to differentiate washout patterns in the training dataset.

Finally, the TCC loss penalizes discrepancies between the image-based classification predictions $\hat{y}^{\mathrm{cls}}_i$ and the physiologically derived labels $\mathrm{label}^{\mathrm{signal}}_i$ across all synthesized phases:
\begin{equation}
    \mathcal{L}_{\mathrm{TCC}} = \sum_{i=1}^{3} \left\| \hat{y}^{\mathrm{cls}}_i - \mathrm{label}^{\mathrm{signal}}_i \right\|^2.
\end{equation}

Here, $f_{\theta}(\cdot)$ explicitly models the physiological contrast trajectory, $g(\cdot)$ generates diagnostic labels consistent with clinical patterns, and $\hat{y}^{\mathrm{cls}}_i$ represents classification predictions derived directly from synthesized image features. Through this constraint, the network is encouraged to generate lesion classification predictions that are both visually accurate and physiologically coherent with the underlying contrast enhancement dynamics.

\subsubsection{Multi-task losses}

To optimize the synthesis of contrast-enhanced MRI, we apply an $\ell_1$ reconstruction loss at each phase, $\mathcal{L}_{\text{syn}} = \sum_{i=1}^{3} \left\| \hat{y}_i - y_i \right\|_1$, where $\hat{y}_t$ is the synthesized image at phase $t$ and $x_t$ is the corresponding ground truth.

For segmentation, we employ a hybrid loss function combining Dice loss and cross-entropy to encourage both overlap and pixel-wise accuracy, $\mathcal{L}_{\text{seg}} = \lambda_{\text{dice}}\, \mathcal{L}_{\text{Dice}}(\hat{y}^{\text{seg}}, y^{\text{seg}}) + \lambda_{\text{ce}}\, \mathcal{L}_{\text{CE}}(\hat{y}^{\text{seg}}, y^{\text{seg}})$, where $\hat{y}^{\text{seg}}$ is the predicted mask, $y^{\text{seg}}$ is the ground truth, and $\lambda_{\text{dice}}, \lambda_{\text{ce}}$ are balancing weights.

For classification, we employ a cross-entropy loss $\mathcal{L}_{\text{cls}} = \mathcal{L}_{\text{CE}}(\hat{y}^{\text{cls}}, y^{\text{cls}})$, where $y^{\text{cls}}$ is the lesion category label.

We empirically set all task-specific loss weights (including $\lambda_{\text{dice}}, \lambda_{\text{ce}}, \lambda_{\text{cls}}, \lambda_{\text{TCC}}$) to 1.0, as this configuration yielded stable optimization and competitive performance across tasks.

\subsection{Experimental Setups}
\subsubsection{Datasets}
We evaluate the performance of our framework on two datasets: the Liver Lesion Diagnosis Challenge on Multi-phase MRI (LLD-MMRI2023)\footnote{https://github.com/LMMMEng/LLD-MMRI2023} and the MG-2021 in-house dataset. 

\begin{figure*}[t]
    \centering
    \includegraphics[width=1\textwidth]{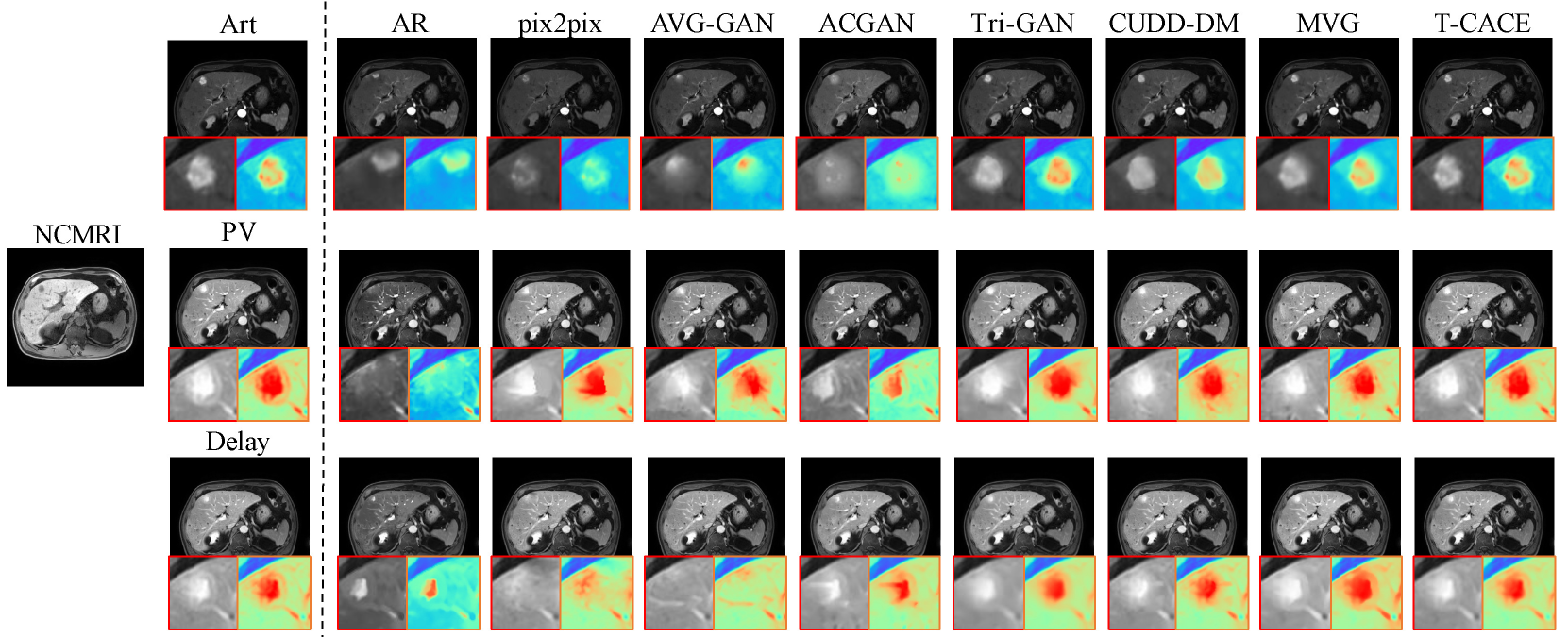}
    \caption{Qualitative comparison of multi-phase contrast-enhanced MRI synthesis results across different methods. Synthesized Art, PV, and Delaye phase images from NCMRI using different methods are shown, with magnified tumor patches and feature maps demonstrating T-CACE's superior structural preservation and contrast evolution.}
    \label{synthesis-1}
\end{figure*}

\textbf{LLD-MMRI2023} consists of 316 training cases, 78 validation cases, and 104 test cases. It provides full-volume data, lesion bounding boxes, and pre-cropped lesion patches. The dataset contains seven different lesion types, including four benign types (hepatic hemangioma, hepatic abscess, hepatic cysts, and focal nodular hyperplasia) and three malignant types (intrahepatic cholangiocarcinoma, liver metastases, and hepatocellular carcinoma). Each lesion is associated with eight different phases (T2WI, DWI, in-phase, out-phase, T1 contrast-pre, contrast-artery, contrast-portal vein, and contrast-delay), providing diverse visual cues. Participants are required to diagnose the type of liver lesion in each case.

\textbf{MG-2021 in-house dataset} comprises 250 subjects who underwent clinically indicated liver MRI examinations at the McGill University Health Centre (MUHC). The dataset includes both pre-contrast sequences—T2 fat-suppressed (T2FS), diffusion-weighted imaging (DWI), and T1-weighted pre-contrast—and contrast-enhanced sequences acquired following gadolinium-based contrast agent (GBCA) administration, including arterial, portal venous, late, and 5-minute delayed phases. All scans were performed on a 3T MRI scanner with a standard GBCA dosage of 0.1 mmol/kg. This retrospective study was approved by the Institutional Review Board (IRB) of MUHC (Approval No. F11HRR-43699), in accordance with local ethics procedures. To ensure anatomical consistency across all phases, manual registration was conducted and verified by three board-certified radiologists with over seven years of diagnostic experience. Lesion type annotations and segmentation masks were reviewed and finalized based on expert consensus to guarantee accurate labeling and reliable downstream analysis.

\subsubsection{Implementation}
Our experiments used a 5-fold cross-validation strategy, with approximately 80\% of the training data and the remaining 20\% for independent testing. We used paired t-tests to evaluate the significance of performance differences between our method and established baselines, with a $p < 0.05$ threshold. The model was implemented in PyTorch and trained on two NVIDIA A100 GPUs under Ubuntu 18.04 using Python 3.6. Training employed the Adam optimizer with an initial learning rate of $10^{-4}$. Owing to the autoregressive nature of our framework, learning rate decay was used to ensure stable phase-wise generation and prevent vanishing gradients in later synthesis steps. Specifically, a cosine annealing schedule was used, with an initial warm-up phase maintaining $10^{-4}$ for the first 20 epochs, followed by linear decay to zero over the remaining epochs. The learning rate was tuned within the range $[10^{-3}, 10^{-4}, 10^{-5}]$ to achieve optimal convergence.

\subsubsection{Evaluation Metrics}
Classification performance was evaluated using accuracy, sensitivity, specificity, and F1-score, which assess the model's ability to distinguish between different lesion types. The quality of synthesized images was evaluated using mean squared error (MSE), peak signal-to-noise ratio (PSNR), structural similarity index (SSIM), learned perceptual image patch similarity (LPIPS), and Fréchet inception distance (FID).  For lesion segmentation, we adopted the Dice similarity coefficient (Dice), Hausdorff distance at 95\% (HD95), and average surface distance (ASD) as quantitative metrics. These segmentation metrics quantify the spatial overlap and boundary accuracy between predicted and reference masks. All these metrics jointly evaluate the fidelity of synthetic CEMRI, the accuracy of lesion delineation, and the overall reliability of the proposed multi-task framework in comparison to ground truth references.

\begin{figure}[t]
	\centering
	\includegraphics[width=0.49\textwidth]{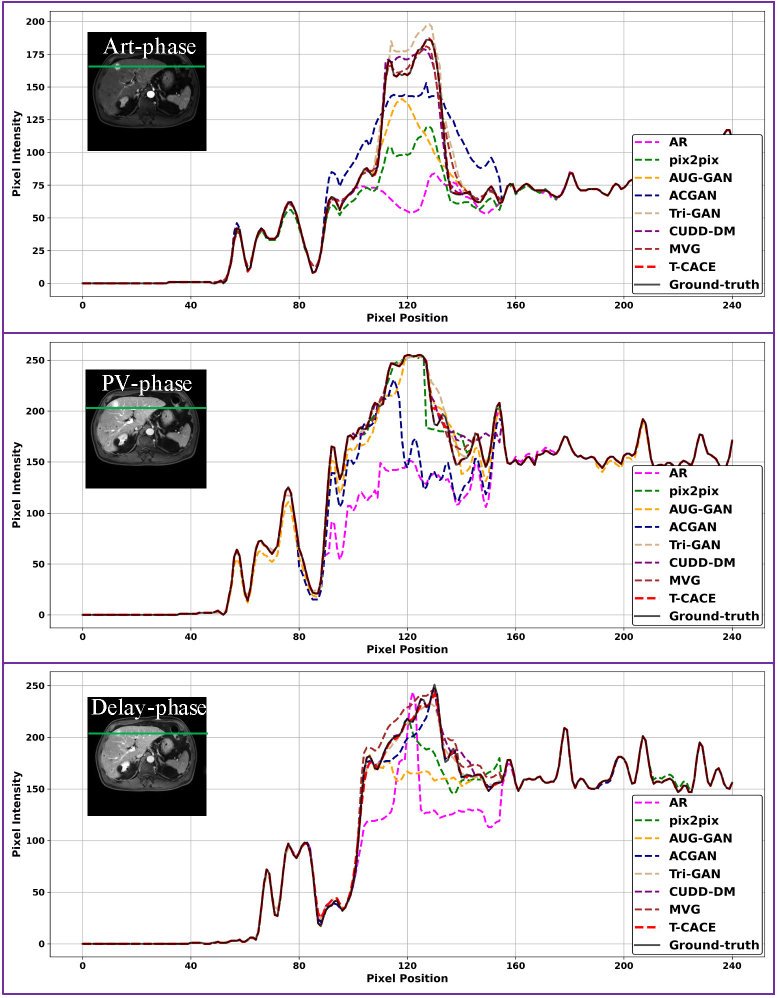}
	\caption{Intensity curves extracted along a horizontal axis ($y=140$) in synthesized images. T-CACE closely matches the ground truth, accurately capturing contrast variations and tumor boundaries, while other methods show artifacts or degraded detail.} \label{synthesis-2}
\end{figure}

\begin{table}[t]
    \centering
    \caption{Quantitative comparison of our method with other methods on the MG-2021 dataset. The synthesis results demonstrate that our method outperforms others in terms of MSE (lower is better, unit: $\times 10^{-2}$), PSNR (higher is better, unit: dB), SSIM (higher is better), LPIPS (lower is better, scaled by $10^{-2}$), and FID (lower is better).}
    \label{mg-synthesis}
    \resizebox{\columnwidth}{!}{ 
    \begin{tabular}{lccccc}
        \toprule
        \textbf{Method} & \textbf{MSE} $\downarrow$ & \textbf{PSNR } $\uparrow$ & \textbf{SSIM} $\uparrow$ & \textbf{LPIPS}  $\downarrow$ & \textbf{FID} $\downarrow$ \\
        \midrule
        \textit{NCMRI → ART} \\
        AR          & 0.527  & 18.52  & 0.706  & 23.17  & 28.15  \\
        pix2pix     & 0.486  & 19.62  & 0.734  & 21.35  & 20.37  \\
        AVG-GAN     & 0.445  & 21.92  & 0.767  & 18.72  & 22.76  \\
        ACGAN       & 0.425  & 22.14  & 0.792  & 18.16  & 22.15  \\
        Tri-GAN     & 0.397  & 23.97  & 0.818  & 17.56  & 19.98  \\
        CUDD-DM     & 0.388  & 24.67  & 0.815  & 16.17  & 18.37  \\
        MVG         & 0.356  & 24.67  & 0.834  & 16.33  & 17.48  \\
        \textbf{T-CACE}  & \textbf{0.327}  & \textbf{25.02}  & \textbf{0.834}  & \textbf{14.92}  & \textbf{17.46}  \\
        \midrule
        \textit{NCMRI → PV} \\
        AR          & 0.523  & 18.57  & 0.709  & 23.19  & 28.11  \\
        pix2pix     & 0.482  & 19.67  & 0.737  & 21.32  & 20.36  \\
        AVG-GAN     & 0.447  & 21.95  & 0.759  & 18.69  & 22.15  \\
        ACGAN       & 0.430  & 23.11  & 0.771  & 18.21  & 21.45  \\
        Tri-GAN     & 0.391  & 24.40  & 0.806  & 17.48  & 20.92  \\
        CUDD-DM     & 0.361  & 24.77  & 0.837  & 16.09  & 18.34  \\
        MVG         & 0.347  & 24.87  & 0.837  & 16.20  & 17.32  \\
        \textbf{T-CACE}  & \textbf{0.312}  & \textbf{25.13}  & \textbf{0.840}  & \textbf{14.93}  & \textbf{17.32}  \\
        \midrule
        \textit{NCMRI → Delay} \\
        AR          & 0.519  & 18.59  & 0.711  & 23.15  & 28.07  \\
        pix2pix     & 0.471  & 19.68  & 0.740  & 21.33  & 26.31  \\
        AVG-GAN     & 0.441  & 21.99  & 0.762  & 18.67  & 22.15  \\
        ACGAN       & 0.427  & 23.11  & 0.773  & 18.08  & 21.12  \\
        Tri-GAN     & 0.391  & 24.12  & 0.806  & 17.42  & 20.92  \\
        CUDD-DM     & 0.383  & 24.48  & 0.819  & 16.11  & 18.71  \\
        MVG         & 0.342  & 24.73  & 0.832  & 16.19  & 18.26  \\
        \textbf{T-CACE}  & \textbf{0.307}  & \textbf{25.19}  & \textbf{0.842}  & \textbf{14.27}  & \textbf{17.26}  \\
        \bottomrule
    \end{tabular}
    }
\end{table}

\begin{table}[t]
    \centering
    \caption{Quantitative comparison of our method with other methods on the LLD-MMRI dataset. The synthesis results demonstrate that our method outperforms others in terms of MSE (lower is better, unit: $\times 10^{-2}$), PSNR (higher is better, unit: dB), SSIM (higher is better), LPIPS (lower is better, scaled by $10^{-2}$), and FID (lower is better).}
    \label{lld-synthesis}
    \resizebox{\columnwidth}{!}{ 
    \begin{tabular}{lccccc}
        \toprule
        \textbf{Method} & \textbf{MSE} $\downarrow$ & \textbf{PSNR} $\uparrow$ & \textbf{SSIM} $\uparrow$ & \textbf{LPIPS} $\downarrow$ & \textbf{FID} $\downarrow$ \\
        \midrule
        \textit{NCMRI → ART} \\
        AR          & 0.468  & 18.74  & 0.712  & 23.02  & 27.54  \\
        pix2pix     & 0.453  & 20.01  & 0.742  & 21.36  & 26.12  \\
        AVG-GAN     & 0.417  & 21.97  & 0.759  & 19.34  & 22.18  \\
        ACGAN       & 0.392  & 23.21  & 0.771  & 18.17  & 22.10  \\
        Tri-GAN     & 0.382  & 24.12  & 0.813  & 17.43  & 20.73  \\
        CUDD-DM     & 0.364  & 24.39  & 0.817  & 16.99  & 18.79  \\
        MVG         & 0.351  & 24.77  & 0.821  & 16.18  & 18.32  \\
        \textbf{T-CACE}  & \textbf{0.323}  & \textbf{25.07}  & \textbf{0.840}  & \textbf{14.35}  & \textbf{17.43}  \\
        \midrule
        \textit{NCMRI → PV} \\
        AR          & 0.460  & 18.80  & 0.715  & 22.89  & 27.48  \\
        pix2pix     & 0.442  & 20.15  & 0.745  & 21.20  & 25.98  \\
        AVG-GAN     & 0.410  & 22.10  & 0.765  & 19.20  & 22.10  \\
        ACGAN       & 0.388  & 23.30  & 0.778  & 18.05  & 21.40  \\
        Tri-GAN     & 0.375  & 24.25  & 0.820  & 17.35  & 20.55  \\
        CUDD-DM     & 0.358  & 24.62  & 0.825  & 16.85  & 18.65  \\
        MVG         & 0.345  & 24.88  & 0.828  & 16.10  & 18.25  \\
        \textbf{T-CACE}  & \textbf{0.319}  & \textbf{25.17}  & \textbf{0.845}  & \textbf{14.20}  & \textbf{17.28}  \\
        \midrule
        \textit{NCMRI → Delay} \\
        AR          & 0.455  & 18.85  & 0.718  & 22.75  & 27.32  \\
        pix2pix     & 0.438  & 20.20  & 0.748  & 21.10  & 25.85  \\
        AVG-GAN     & 0.405  & 22.20  & 0.770  & 19.10  & 22.00  \\
        ACGAN       & 0.385  & 23.40  & 0.780  & 18.00  & 21.20  \\
        Tri-GAN     & 0.370  & 24.30  & 0.825  & 17.25  & 20.50  \\
        CUDD-DM     & 0.355  & 24.70  & 0.830  & 16.75  & 18.55  \\
        MVG         & 0.342  & 24.98  & 0.835  & 16.05  & 18.15  \\
        \textbf{T-CACE}  & \textbf{0.315}  & \textbf{25.25}  & \textbf{0.850}  & \textbf{14.05}  & \textbf{17.15}  \\
        \bottomrule
    \end{tabular}
    }
\end{table}

\subsection{Multi-phase CEMRI Synthesis Evaluation}

\subsubsection{Qualitative analysis and comparative evaluation} To comprehensively assess the performance of our proposed method, we compare T-CACE against seven existing methods—Autoregressive models (AR) \cite{van2016pixel}, pix2pix \cite{Huang2006}, AUG-GAN \cite{frid2018}, ACGAN \cite{frid2018gan}, Tri-GAN \cite{zhao2020tripartite}, CUDD-DM \cite{xu2024common}, and MVG \cite{ren2024medical}—on two independent datasets: MG-2021 and LLD-MMRI.

Fig.~\ref{synthesis-1} presents visual comparisons for synthesizing three contrast-enhanced MRI phases (Art, PC, and Delay) from non-contrast MRI (NCMRI). Enlarged local patches of tumor regions and corresponding feature maps are shown alongside the synthesized images to facilitate intuitive comparison. Qualitative analysis demonstrates that T-CACE achieves superior synthesis performance, producing results that closely approximate the ground truth images. Closer inspection of magnified regions reveals that T-CACE effectively captures fine tumor features and preserves critical structural details, enabling realistic modeling of contrast transitions. In contrast, baseline methods often exhibit deficiencies such as loss of tumor information, poor rendering of fine textures, and blurred lesion boundaries. These shortcomings are particularly pronounced in regions with ambiguous lesion margins or low lesion-to-parenchyma contrast. The superior performance of T-CACE can be attributed to its time-conditioned autoregressive synthesis strategy, which leverages sequential dependencies between phases to ensure physiologically coherent contrast evolution.

To further assess the synthesis quality of different methods, we analyze pixel intensity profiles along a horizontal line at \( y = 140 \), as illustrated in Fig.~\ref{synthesis-2}. Each curve represents the pixel intensities sampled from the same row across synthesized images from all competing methods, with the ground truth profile serving as reference. We focus particularly on the region between \( x = 80 \) and \( x = 155 \), which corresponds to the lesion area and contains significant intensity variations that reflect underlying anatomical boundaries. In this region, the curve produced by T-CACE demonstrates remarkable consistency with the ground truth, precisely reproducing the positions and amplitudes of both peaks and valleys. This highlights T-CACE's strong capability in preserving subtle intensity transitions and fine-grained lesion structure. In contrast, baseline methods exhibit noticeable deviations. Specifically, AR severely underestimates signal intensity across the lesion region (\( x = 110 \) to \( x = 155 \)), resulting in attenuated contrast and reduced delineation of lesion structures. pix2pix and ACGAN curves show systematic errors, such as intensity undershooting or overly smoothed transitions, which result in the blurring of lesion boundaries. These discrepancies suggest that many baselines struggle to capture detailed anatomical characteristics within complex pathological regions. Overall, the intensity profile visualization demonstrates T-CACE’s superior fidelity in modeling tumor-specific signal dynamics.

\subsubsection{Quantitative Comparison with Existing Methods}

Quantitative results demonstrate that T-CACE consistently outperforms existing methods across all three synthesis tasks (\textit{NCMRI $\rightarrow$ ART}, \textit{NCMRI $\rightarrow$ PV}, and \textit{NCMRI $\rightarrow$ Delay}). Specifically, our method achieves the lowest MSE, highest PSNR, highest SSIM, lowest LPIPS, and lowest FID across all phases, confirming its capability to synthesize high-fidelity contrast-enhanced MRI images. 

On the MG-2021 dataset (Table~\ref{mg-synthesis}), T-CACE achieves an MSE of 0.307, a PSNR of 25.19 dB, an SSIM of 0.842, an LPIPS of $14.27 \times 10^{-2}$, and a FID of 17.36 in the \textit{NCMRI $\rightarrow$ Delay} synthesis task. Compared to the second-best method (MVG), T-CACE improves PSNR by 0.46 dB and SSIM by 0.01, highlighting its superior contrast preservation and structural consistency. This advantage is further validated on the LLD-MMRI dataset (Table~\ref{lld-synthesis}). Although the overall synthesis performance improves for all methods with the increased dataset size, T-CACE consistently remains the best-performing model.

These results demonstrate that our autoregressive framework effectively captures the progressive nature of contrast enhancement while preserving anatomical integrity. The proposed sequential synthesis strategy ensures that each contrast phase is generated in order, leveraging information from preceding phases to reinforce structural consistency and produce realistic contrast distributions. Thanks to its time-conditioned autoregressive design and dynamic time-aware attention mechanism, T-CACE consistently delivers stable and superior performance across all synthesis tasks. The alignment between quantitative metrics and qualitative observations further underscores the strong synthesis capabilities of our proposed model.

\begin{figure*}[t]
	\centering
	\includegraphics[width=1\textwidth]{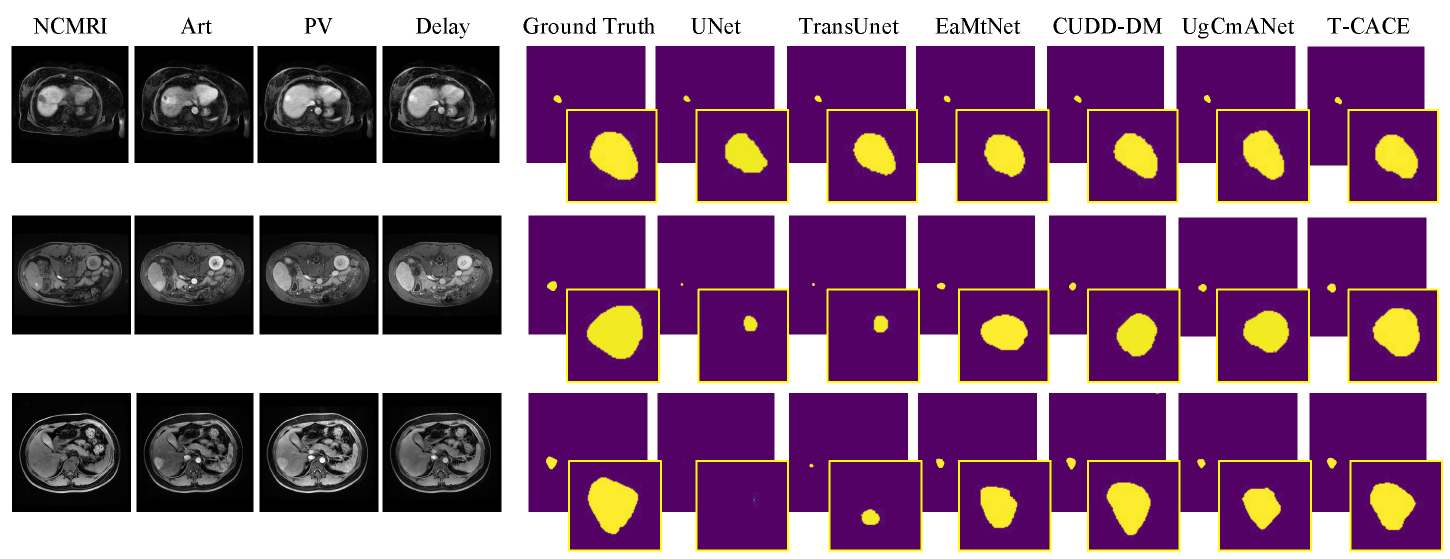}
	\caption{Visualization of segmentation results produced by six models on representative samples. Ground truth (GT) and predictions are displayed as colored masks. The cropped region highlights the lesion area for clearer boundary comparison.} \label{seg}
\end{figure*}

\subsection{Lesion Segmentation Evaluation}
\subsubsection{Qualitative Segmentation Results and Discussion}

To qualitatively evaluate the effectiveness of our proposed method, Fig.~\ref{seg} presents visual comparisons of segmentation results, including ground truth (GT) masks and predictions from five existing approaches alongside our own. For improved interpretability, all binary masks are visualized as color maps, with consistent lesion-centered cropping across all examples. Specifically, UNet~\cite{ronneberger2015u} and TransUNet~\cite{chen2021transunet} rely solely on non-contrast-enhanced images, which lack vascular enhancement, making accurate tumor boundary delineation challenging. Their segmentation outputs exhibit noticeable under-segmentation and boundary deviations, particularly in low-contrast regions. Although TransUNet provides minor improvements over UNet by incorporating global attention mechanisms, it still struggles to achieve complete lesion coverage. EaMtNet~\cite{xiao2023edge} and UgCmA-Net~\cite{zhao2025uncertainty} incorporate multi-phase contrast-enhanced information, allowing for improved structural delineation. However, their performance is still constrained by inter-phase misalignment and suboptimal fusion strategies, which can introduce boundary noise and occasional false positives. CUDD-DM~\cite{xu2024common} adopts a synthesis-then-segmentation strategy by generating pseudo-contrast images for downstream segmentation. While this approach enhances lesion visibility, the segmentation results are affected by artifact-driven errors and reduced boundary precision due to imperfections in the synthesized modalities.

In contrast, the proposed T-CACE framework jointly optimizes synthesis and segmentation under a unified time-conditioned autoregressive paradigm, reinforced by temporal coherence and cross-modal constraints. Consequently, our model consistently produces superior segmentation outputs, achieving accurate boundary delineation, complete lesion coverage, and minimal false positives, closely approximating the ground truth even in complex and challenging cases.

\subsubsection{Quantitative Evaluation of Segmentation Results and Discussion}

\begin{figure}[t]
	\centering
	\includegraphics[width=0.49\textwidth]{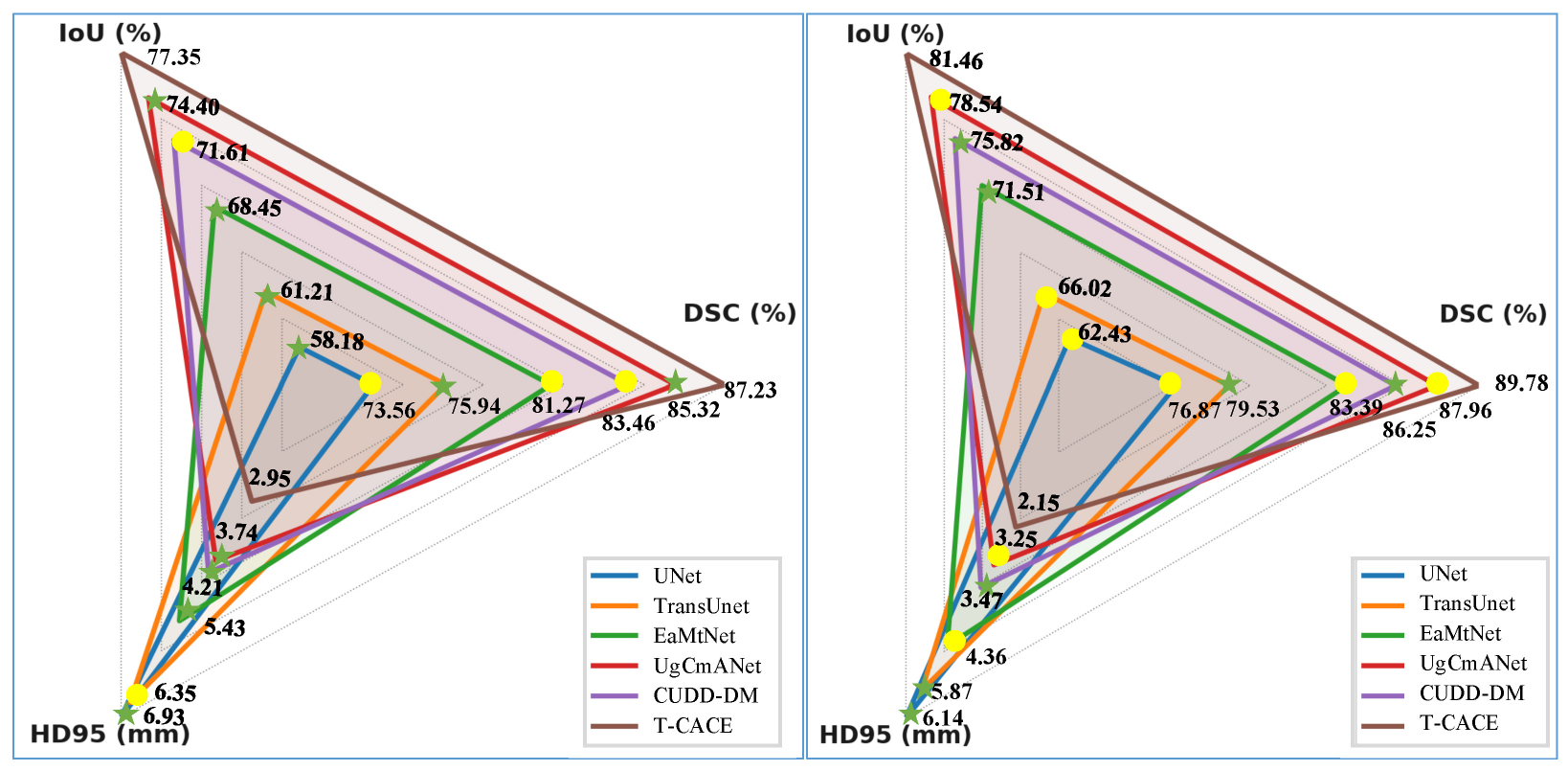}
	\caption{Triangle radar plot comparison of six segmentation methods across three metrics (DSC, IoU, HD95) on the MG and LLD datasets. Statistical significance is annotated with green stars (\textcolor{green}{\ding{72}}, $p < 0.05$) and yellow solid circles (\textcolor{yellow}{\ding{108}}), $p < 0.01$), based on paired t-tests over five cross-validation folds.} \label{radar}
\end{figure}

To validate the effectiveness of our proposed method, T-CACE, we conducted comparative experiments against five widely used segmentation models: UNet~\cite{ronneberger2015u}, TransUNet~\cite{chen2021transunet}, EaMtNet~\cite{xiao2023edge}, CUDD-DM~\cite{xu2024common}, and UgCmA-Net~\cite{zhao2025uncertainty}. All methods were evaluated on two public liver MRI datasets (MG-2021 and LLD-MMRI) using three standard performance metrics: Dice similarity coefficient (DSC), intersection over union (IoU), and 95th percentile Hausdorff distance (HD95). As illustrated in Fig.~\ref{radar}, T-CACE consistently outperformed all baselines across all three metrics on both datasets. Specifically, T-CACE achieved the highest DSC and IoU scores while maintaining the lowest HD95 value, indicating both superior overlap accuracy and more precise boundary delineation. 

To further assess the statistical robustness of the observed improvements, we performed \textit{paired t-tests} across the five folds of cross-validation. Statistical significance is annotated directly on the radar plots. For example, T-CACE achieved statistically significant improvements in DSC compared to EaMtNet and CUDD-DM ($p < 0.01$), and in IoU compared to UNet and MVG ($p < 0.05$). These results confirm that the performance gains of T-CACE are statistically meaningful rather than incidental. Notably, the performance margins were particularly pronounced on the LLD-MMRI dataset, where baseline methods exhibited larger HD95 values, suggesting weaker boundary precision in more complex cases. In contrast, T-CACE achieved both the best HD95 and significant improvements in DSC and IoU, demonstrating its superior generalization ability across datasets with diverse liver lesion characteristics and imaging protocols.

\begin{figure*}[t]
	\centering
	\includegraphics[width=1\textwidth]{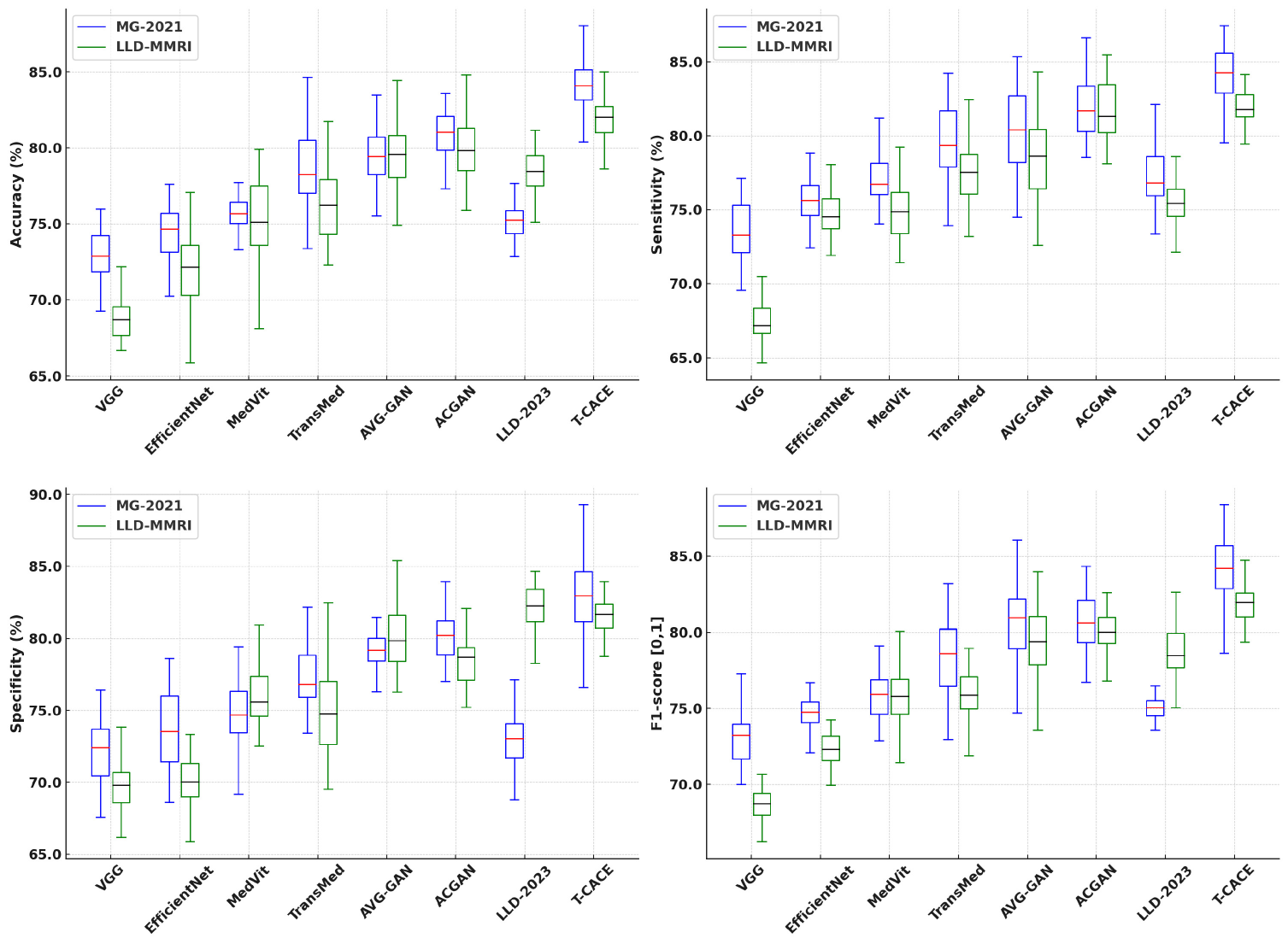}
	\caption{Boxplot comparison of tumor classification performance between our method and other SOTA methods. Classification accuracy, sensitivity, specificity, and F1-score of different methods on the MG and LLD datasets. T-CACE achieves consistently higher and more stable performance.} \label{classification}
\end{figure*}

\subsection{Lesion Classification Evaluation}

To quantitatively evaluate the classification performance of our method, we compare T-CACE against several existing approaches, including VGG~\cite{simonyan2014very}, EfficientNet~\cite{tan2019efficientnet}, MedViT~\cite{manzari2023medvit}, TransMed~\cite{dai2021transmed}, AUG-GAN~\cite{frid2018}, ACGAN~\cite{frid2018gan}, and LLD-2023\footnote{https://github.com/ZHEGG/miccai2023}, on both the MG-2021 and LLD-MMRI datasets. As illustrated in Fig.~\ref{classification}, T-CACE consistently achieves the highest classification performance across all evaluation metrics.

These results demonstrate that T-CACE effectively captures lesion-specific features, leading to improved discrimination between different lesion types. Compared to conventional classification models that directly operate on non-contrast MRI (NCMRI), T-CACE benefits from synthesized contrast-enhanced representations, enabling the model to learn richer diagnostic patterns. A key observation is that methods incorporating synthesized contrast-enhanced images (e.g., AUG-GAN, ACGAN, and our proposed T-CACE) generally outperform those directly classifying NCMRI (e.g., VGG, EfficientNet, TransMed, MedViT). This performance gain can be attributed to the enhanced visual contrast and improved lesion visibility provided by the generative models, which facilitate more accurate tumor boundary delineation and internal texture characterization for the downstream classifiers.

Notably, our T-CACE framework achieves the best overall classification performance among all methods, owing to its progressive phase-aware synthesis strategy and joint optimization of synthesis and classification tasks. Although the LLD-2023 model achieved competitive results in its original report, its performance is lower in our evaluation, likely due to differences in training settings—our experiments are conducted entirely under the non-contrast condition, while LLD-2023 was originally trained with real CEMRI data. Overall, these findings validate the effectiveness and generalizability of T-CACE in improving lesion classification performance by leveraging autoregressive synthesis of contrast-enhanced images, offering a promising non-invasive alternative for liver lesion diagnosis.


\begin{table*}[t]
\centering
\caption{Ablation study of different components in the T-CACE framework on the MG-2021 dataset. Synthesis, segmentation, and classification performance are reported.}
\label{ablation}
\begin{tabular}{lcccccccccccc}
\toprule
\textbf{Variant} & \multicolumn{5}{c}{\textbf{Synthesis Metrics}} & \multicolumn{3}{c}{\textbf{Segmentation Metrics}} & \multicolumn{4}{c}{\textbf{Classification Metrics}} \\
\cmidrule(lr){2-6} \cmidrule(lr){7-9} \cmidrule(lr){10-13}
 & MSE  & PSNR  & SSIM  & LPIPS  & FID  & DSC  & IoU  & HD95 & Accuracy & Sensitivity & Specificity & F1-score \\
\midrule
Baseline     & 0.519 & 18.59 & 0.711 & 23.15 & 28.07 & 74.91 & 62.05 & 6.31 & 72.81 & 71.67 & 71.94 & 72.79 \\
No DTAM      & 0.328 & 25.43 & 0.831 & 15.79 & 18.86 & 85.42 & 75.98 & 3.62 & 77.10 & 78.62 & 76.58 & 77.02 \\
No CTE       & 0.320 & 25.34 & 0.828 & 15.88 & 18.65 & 85.60 & 76.21 & 3.41 & 78.50 & 79.37 & 77.63 & 78.50  \\
No T-Encoding& 0.319 & 25.72 & 0.837 & 15.17 & 18.35 & 86.51 & 76.94 & 3.05 & 80.82 & 81.48 & 80.16 & 80.81\\
\textbf{T-CACE} & \textbf{0.307} & \textbf{25.19} & \textbf{0.842} & \textbf{14.27} & \textbf{17.26} & \textbf{87.23} & \textbf{77.35} & \textbf{2.93} & \textbf{83.93} & \textbf{84.21} & \textbf{83.64} & \textbf{83.92} \\
\bottomrule
\end{tabular}
\end{table*}

\subsection{Ablation study}
\subsubsection{Quantitative Analysis of Module Effectiveness}
To comprehensively evaluate the contribution of each key component in the proposed T-CACE (Time-Conditioned Autoregressive Contrast Enhancement) framework, we conducted an ablation study using the delayed-phase contrast-enhanced MRI (Delay) synthesis task from non-contrast MRI $x_{NC}$ as a representative setting. Specifically, we first define a baseline model (\textbf{Baseline}), where all temporal and structural conditioning mechanisms are removed, and the autoregressive generator operates directly on raw $x_{NC}$ without any auxiliary tokens or priors. To investigate the impact of temporal modeling, two additional ablations are performed: removing the continuous time embedding (\textbf{No T-Encoding}) and disabling the Gaussian-decayed dynamic attention mask (\textbf{No DTAM}). Furthermore, the importance of anatomical priors is assessed by omitting the conditional token encoding (\textbf{No CTE}). The complete model (\textbf{T-CACE}) integrates all components without ablations.

As shown in Table~\ref{ablation}, removing any module results in a consistent decline in synthesis fidelity, segmentation accuracy, and classification performance. Notably, the ablation of DTAM leads to significant reductions in SSIM (-0.011) and Dice coefficient (-1.81), highlighting the critical role of temporal attention in preserving spatial structure and lesion boundaries. Additionally, classification performance is notably impaired when DTAM is removed, with accuracy decreasing by 6.83\% and F1-score declining by 6.90\%. Omitting CTE also leads to noticeable drops in classification metrics (accuracy -5.43\%, F1-score -5.42\%), underscoring the importance of anatomical priors for diagnostic consistency. Furthermore, excluding the temporal encoding component (``No T-Encoding'') results in moderate but clear performance degradation across all evaluated metrics, confirming its complementary contribution to maintaining temporal coherence. These results collectively confirm that each module provides complementary and essential contributions to the overall performance. Their joint integration is crucial for achieving robust, anatomically consistent, temporally coherent, and diagnostically reliable multi-phase CEMRI synthesis, segmentation, and classification.

\subsubsection{Qualitative Analysis of Module Effectiveness}

\begin{figure}[t]
	\centering
	\includegraphics[width=0.48\textwidth]{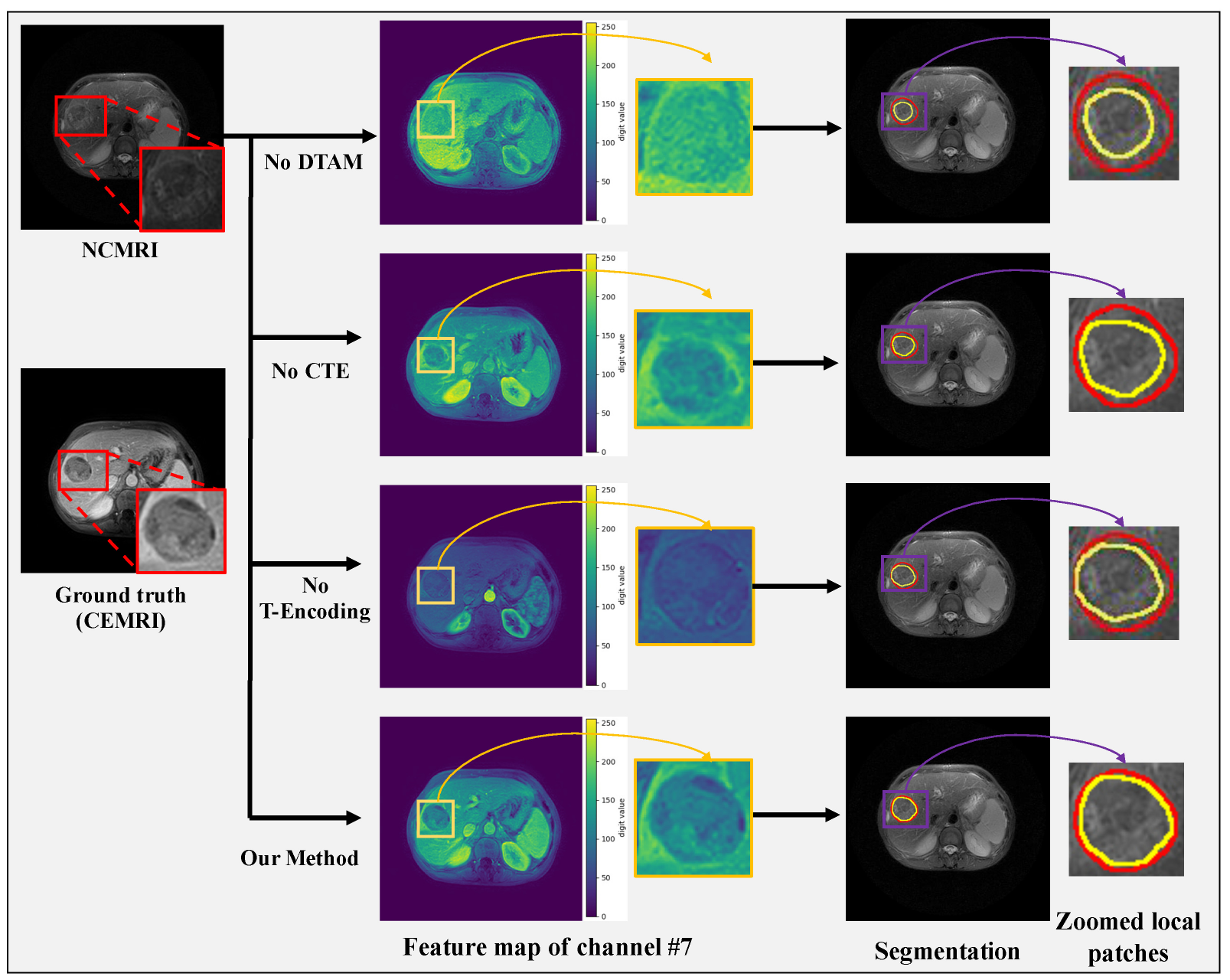}
	\caption{Qualitative comparison of feature responses and segmentation results under different ablation settings. From top to bottom: No DTAM, No CTE, No T-Encoding, and the full model (``T-CACE''). The middle column displays the activation map from feature channel \#7, while the right column shows segmentation outputs with zoomed-in tumor contours. Yellow and red contours denote the predicted segmentation and ground truth, respectively. These visualizations highlight the impact of each component on spatial encoding, temporal coherence, and lesion boundary delineation. } \label{fig:ablation1}
\end{figure}

To further validate the contributions of individual components within the proposed T-CACE framework, we present a qualitative comparison of feature representations and segmentation results under different ablation settings, as shown in Figure~\ref{fig:ablation1}. Specifically, we visualize the activation maps from feature channel~7, the corresponding segmentation outputs, and zoomed-in tumor regions for four model variants: ``No DTAM,'' ``No CTE,'' ``No T-Encoding,'' and the complete model (``T-CACE''). These visual comparisons are consistent with the quantitative trends reported in Table~\ref{ablation}, and offer intuitive insights into how each module contributes to the overall performance. In particular, removing CTE leads to degraded boundary sharpness and reduced lesion contrast, indicating the importance of anatomical priors in enhancing spatial discriminability. When DTAM is ablated, the feature activations become less localized and temporally inconsistent, suggesting weakened phase-wise attention and impaired temporal modeling. Similarly, removing T-Encoding disrupts the model’s ability to encode phase-specific dynamics, resulting in blurred feature responses and reduced phase contrast. In contrast, the full T-CACE model produces more distinct feature activations and sharper segmentation contours, especially in cases with small or poorly contrasted lesions. These results highlight that each module plays a distinct yet complementary role—CTE for anatomical structure preservation, and DTAM and T-Encoding for phase-aware temporal consistency—jointly enabling high-fidelity lesion representation and accurate downstream segmentation in synthesized contrast-enhanced MRI.

\section{Conclusion and Future work}
This work presents T-CACE, a unified time-conditioned autoregressive framework that enables contrast-free liver MRI synthesis, segmentation, and classification. By integrating Conditional Token Encoding (CTE), the model effectively incorporates anatomical and temporal priors. The Dynamic Time-aware Attention Mask (DTAM) ensures physiologically coherent enhancement across sequential phases, while the Temporal Classification Consistency (TCC) constraint aligns diagnostic predictions with underlying signal dynamics. Extensive experiments on two liver MRI datasets demonstrate that T-CACE achieves consistent improvements over state-of-the-art methods. These results highlight the potential of T-CACE as a safe and effective alternative to traditional contrast-enhanced imaging, offering a reliable solution for contrast-free liver disease assessment.

Although the proposed T-CACE framework demonstrates strong performance across two abdominal MRI datasets, several limitations remain that warrant further investigation. First, while our model leverages continuous-time embeddings to handle variations in interphase intervals flexibly, its robustness under highly non-standard or irregular acquisition protocols has not been explicitly validated. In future work, we plan to systematically evaluate the framework on external multi-center cohorts with diverse temporal sampling schemes to establish its generalizability further.  Second, although this study focuses on liver tumor synthesis and classification, the T-CACE framework is designed to be modular and organ-agnostic. Both Conditional Token Encoding (CTE) and Dynamic Time-Aware Attention Mask (DTAM) modules are adaptable to new anatomical contexts, such as the kidney or pancreas, through data-driven learning. Nonetheless, since contrast enhancement dynamics may vary across organs (e.g., more heterogeneous uptake in the pancreas), additional refinements such as soft anatomical priors or organ-specific time encodings may improve performance. Future work will investigate these extensions and validate the model's applicability across broader multi-organ clinical settings. Finally, our current design does not explicitly estimate uncertainty across tasks such as image synthesis or tumor classification. Incorporating uncertainty-aware mechanisms (e.g. evidential modeling or Bayesian neural networks) may help quantify prediction confidence, identify ambiguous regions, and improve clinical trust. Future work will explore the integration of such techniques into the T-CACE framework to enable risk-aware modeling and selective refinement strategies.

\balance
\section{REFERENCES}
\bibliographystyle{IEEEtran}
\bibliography{mybib}

\end{document}